# A test for partial correlation with censored astronomical data

M.G. Akritas[1] and J.Siebert[2]

[1] *Department of Statistics, Pennsylvania State University, State College PA 16802*

[2] *Max–Planck–Institut für Extraterrestrische Physik, 85740 Garching, Germany*

**ABSTRACT**

A new procedure is presented, which allows, based on Kendall's $\tau$, to test for partial correlation in the presence of censored data. Further, a significance level can be assigned to the partial correlation – a problem which hasn't been addressed in the past, even for uncensored data. The results of various tests with simulated data are reported. Finally, we apply this newly developed methodology to estimate the influence of selection effects on the correlation between the soft X–ray luminosity and both total and core radio luminosity in a complete sample of Active Galactic Nuclei.

## 1 INTRODUCTION

Astronomers are frequently confronted with the problem of missing or incomplete information. This typically happens when a sample of sources, which has been selected for showing emission in a certain waveband, is then observed in another part of the electromagnetic spectrum. A lack of intrinsic emission, absorption due to intervening material or insufficient sensitivity of the instrument then often result in upper limits or, more general, in a 'censored' data set.

In our specific case a sample of Active Galactic Nuclei (AGN) with 2.7 GHz fluxes greater than 2 Jy has been established for which almost complete information on the radio and the optical continuum as well as line emission exists (Morganti et al. 1993, Tadhunter et al. 1993, di Serego Alighieri et al. 1994). The soft X–ray properties of these objects were determined by using the flux limited ROSAT All–Sky Survey (Siebert et al. 1995). For about 40% of the objects in this sample only an upper limit on the soft X–ray flux could be given.

One probable clue to the radiation mechanisms in AGN is to search for a relationship between the emission from different wavebands. Many attempts have been made in the past to investigate the correlations between the radio, the optical and the X–ray regime (e.g. Feigelson & Berg 1983, Fabbiano et al. 1984, Zamorani 1984, Kembhavi et al. 1986, Wilkes & Elvis 1987, Browne & Murphy 1987). Thus, given the above mentioned problems, many procedures have been developed by astrophysicists to deal with the problem of correlation and regression analysis with censored data (Schmitt 1985, Feigelson & Nelson 1985, Isobe et al. 1986, Avni & Tananbaum 1986).

By applying regression and correlation analysis to the radio and soft X–ray continuum emission of the above mentioned complete sample including the upper limits (using ASURV Rev 1.3, La Valley et al. 1992, Feigelson & Nelson 1985, Isobe et al. 1986), correlations of the soft X–ray luminosity with both the total and the radio core luminosity were found (Siebert et al. 1995). The use of luminosities instead of fluxes, however, always introduces a redshift bias to the data, as luminosities are strongly correlated with redshift in flux limited samples. It is therefore crucial to estimate the influence of this effect on the correlations in order to be able to draw reliable conclusions on the true physical relationship between the emission from the two wavebands. Partial correlation coefficients have been used to deal with this problem (e.g. Kembhavi et al. 1986). However, up to now, censored data could not be taken into account.

In this paper we want to present a method that allows to apply partial correlation to censored data and to assign a significance level to the resulting correlation coefficient. The structure of the paper is as follows: after introducing the notation and partial Kendall's $\tau$ coefficient (§2.1), this concept will be extended to censored data (§2.2). In §3 we describe the various tests we applied and report numerical results on both simulated and 'real' data.

We note that our method is based on rank correlation coefficients. Rank correlation analysis is more general than the frequently used linear correlation analysis and thus our method is also applicable to linear correlation coefficients.

This procedure resulted from a interdisciplinary collaboration of astrophysics and mathematical statistics in the form of the newly founded *Statistical Consulting Center for Astronomy* (SCCA). Further infor-mation can be obtained through World Wide Web (http://www.stat.psu.edu/scca/homepage.html), or by contacting SCCA@stat.psu.edu. The computer code developed on the basis of the procedure presented in this paper is also available from this site.

## 2 PARTIAL KENDALL'S $\tau$ COEFFICIENT WITH CENSORED DATA

In this section we give a description of the partial Kendall's $\tau$ coefficient with censored data and describe a procedure for testing the hypothesis that the population partial Kendall's $\tau$ is zero. In the first subsection we give a brief introduction and background references for Kendall's rank correlation coefficient and Kendall's partial rank correlation coefficient with uncensored data. The procedure for censored data is given in subsection 2.2.

### 2.1 Introduction and Background

In this subsection we consider the uncensored case. Let $\mathbf{T} = (T_1, T_2, T_3)$ be the random vector of interest, and let $\mathbf{T}_i = (T_{1i}, T_{2i}, T_{3i})$, $i = 1, \ldots, n$, be the sample values. For $k = 1, 2, 3$, set

$$J_k(i,j) = I(T_{ki} < T_{kj}) - I(T_{kj} < T_{ki}),$$

where $I(x < y) = 1$, if $x < y$ and 0 otherwise. Kendall's (1938) rank correlation coefficient between $T_k$ and $T_l$ is defined by

$$\tau_{kl} = E(J_k(i,j)J_l(i,j)), \;\; k \neq l,$$

and its sample estimate by

$$\hat{\tau}_{kl} = \frac{2}{n(n-1)} \sum_{i<j} J_k(i,j)J_l(i,j).$$

It has been shown that $\tau$ can be extended to the case of partial correlation and that the partial $\tau$ has the same structural form as $\rho_{12.3}$, the Pearson's partial product-moment correlation (Kendall 1970). In particular, Kendall's partial rank correlation coefficient between $T_1$ and $T_2$ given $T_3$ is defined as

$$\tau_{12.3} = \frac{\tau_{12} - \tau_{13}\tau_{23}}{[(1-\tau_{13}^2)(1-\tau_{23}^2)]^{1/2}}.$$

For a general discussion of the problem of measuring partial association see Quade (1974). A geometric interpretation of partial correlation is given in Thomas & O'Quigley (1993). In spite of the long history of Kendall's partial rank correlation coefficient, there are no tests for the significance of the partial $\tau$ (Hettmansperger 1984). See also Nelson & Yang (1988) where they study, via Monte Carlo, the performance of the Jackknife approximation to the distribution of $\hat{\tau}_{12.3}$. A useful discussion on the interpretation of Kendall's partial rank correlation coefficient can also be found in Nelson & Yang (1988).

### 2.2 Extension to Censored Data

The extension of Kendall's $\tau$ to censored data was first given by Brown, Hollander & Korwar (1974) in a biostatistical context. A more careful derivation of its distributional properties was given by Oakes (1982). After introducing some notation, we describe this censored data version of Kendall's $\tau$. The partial $\tau$ is then defined in terms of $\tau$ as in the uncensored case. Then we describe a method for testing the significance of the partial $\tau$. To our knowledge, this method is new even with uncensored data, since Macklin (1982) only verified by computer simulations that the asymptotic distribution of Spearman's partial $\rho$ has, under the null hypothesis, the same form as the asymptotic distribution of Spearman's $\rho$.

Let again $\mathbf{T} = (T_1, T_2, T_3)$ be the random vector of interest. However, due to censoring we only observe $(X_{1i}, \delta_{1i}, X_{2i}, \delta_{2i}, X_{3i}, \delta_{3i})$, $i = 1, \ldots, n$, where, for $k = 1, 2, 3$, $X_{ki} = min\{T_{ki}, C_{ki}\}$, $\delta_{ki} = I(T_{ki} \leq C_{ki})$ where $C_{ki}$ is the censoring variable and $I(A)$ is the indicator of the event $A$.

At this point we have to emphasize that the above is the right censoring model common in Biostatistics. In Astronomy the data are generally left censored. Left censoring, however, can be converted to right censoring by multiplying all data points by $-1$. (If the log of the data is being analyzed, multiplication by $-1$ should take place after taking logs.) With this conversion, $C_{ki}$ represents minus (the log of) the detection limit for the $k-th$ coordinate of the $i-th$ observation, $T_{ki}$ is minus (the log of) the $k-th$ coordinate of the $i-th$ observation, and if $\delta_{ki} = 1$ then what is observed (i.e. $X_{ki}$) is the variable of interest, while if $\delta_{ki} = 0$ then only the detection limit was recorded.

The censored data version of the function $J$ becomes

$$J_k(i,j) = \delta_{ki} I(X_{ki} < X_{kj}) - \delta_{kj} I(X_{kj} < X_{ki}).$$

For $k, \; l = 1, 2, 3$, set

$$h_{kl}(i,j) = J_k(i,j)J_l(i,j).$$

In this notation, the censored data version of Kendall's $\tau$ between $T_k$ and $T_l$ is

$$\hat{\tau}_{kl} = \frac{2}{n(n-1)} \sum_{i<j} h_{kl}(i,j),$$

and the censored data version of the partial Kendall's $\tau$ between $T_1$ and $T_2$ given $T_3$ is

$$\hat{\tau}_{12.3} = \frac{\hat{\tau}_{12} - \hat{\tau}_{13}\hat{\tau}_{23}}{[(1-\hat{\tau}_{13}^2)(1-\hat{\tau}_{23}^2)]^{1/2}}.$$

Under the null hypothesis $H_0$ that the partial Kendall's $\tau$ is zero, the above statistic is asymptotically normal with zero mean and estimated variance (see also appendix)

$$\hat{\sigma}^2 = 16n^{-1} \frac{A_n}{(1-\hat{\tau}_{13}^2)(1-\hat{\tau}_{23}^2)},$$

where

$$A_n = (n-1)^{-1} \sum_{i_1=1}^{n} \left(B_{i_1} - \bar{B}\right)^2, \qquad (1)$$

where

$$B_{i_1} = \frac{6}{(n-1)(n-2)(n-3)} \sum_{\substack{j_1 < i_2 < j_2 \\ \text{all } \neq i_1}} g(i_1, j_1, i_2, j_2),$$

$\bar{B}$ is the average of the $B_i$'s, and

$$g(i_1,j_1,i_2,j_2) = \frac{1}{24}\sum_p \tilde{g}(i_1,j_1,i_2,j_2)$$

where $\sum_p$ denotes summation over all permutations of $(i_1, j_1, i_2, j_2)$ and

$$\tilde{g}(i_1,j_1,i_2,j_2) = h_{12}(i_1,j_1) - h_{13}(i_1,j_1)h_{23}(i_2,j_2).$$

The hypothesis of zero partial correlation coefficient is rejected at level $\alpha$ if

$$|\frac{\hat{\tau}_{12.3}}{\hat{\sigma}}| > z_{\alpha/2},$$

where $z_{\alpha/2}$ denotes the $100(1-\alpha/2)$-th percentile of the standard normal distribution.

## 3 NUMERICAL RESULTS AND DATA ANALYSIS

### 3.1 Simulations

The testing procedure described in Section 2 is based on the asymptotic (i.e. 'large' sample size) normality of the partial $\tau$. In practice, however, we often have to deal with small or moderate sample sizes. Thus it is useful to have some understanding of the performance of the procedure under such settings. Two important performance characteristics of any testing procedure are the *attained level* and the *power* of the procedure. With finite samples, the attained level will not be exactly equal to the chosen $\alpha$ because the small-sample distribution of the test statistic is not exactly normal. The power of a testing procedure is the probability that the procedure will reject the null hypothesis when it is not true. Clearly, the more pronounced the departure from the null hypothesis, the greater the power.

Both the attained level and the power of a testing procedure against selected alternatives can be evaluated via simulation studies using artificially generated data sets. For the simulation results reported we used sample size $n = 30$, and $\alpha = 0.05$. Under the null hypothesis (i.e. zero partial correlation) the data sets were generated as follows: $T_{1i}$, $T_{2i}$, $T_{3i}$ are all independent exponential random variables with mean one; since all variables are generated independently, the partial correlation coefficient between $T_1$ and $T_2$ given $T_3$ is zero. The censoring variables $C_{1i}$, $C_{2i}$, $C_{3i}$ are independent exponential random variables with mean four. This gives a theoretical level of censoring of 20% for all three variables. The statistic was based on the data $X_{ki} = min\{T_{ki}, C_{ki}\}$, $\delta_{ki} = I(T_{ki} \leq C_{ki})$, $k = 1,2,3$, $i = 1,\ldots,30$. From 1000 simulated data sets the null hypothesis was rejected 71 times. Next, in order to get an idea of how sensitive the test is to departures from the null hypothesis, random samples were generated with nonzero partial correlation. Four levels of departure from the null hypothesis were considered. For all levels the variable $T_{3i}$ and all the censoring variables were generated as before. Variables $T_{1i}$ and $T_{2i}$ were generated as follows: For the first level, $T_{1i} = 0.8T^*_{1i} + 0.2T_{4i}$, $T_{2i} = 0.8T^*_{2i} + 0.2T_{4i}$, where $T^*_{1i}$, $T^*_{2i}$, $T_{4i}$ are all independent exponential random variables (and independent from $T_{3i}$) with mean one. Thus $T_{1i}$, $T_{2i}$ are dependent due to the presence of the common $T_{4i}$ and this dependence is the same when the independent $T_{3i}$ is held fixed. For the second level, $T_{1i} = 0.6T^*_{1i} + 0.4T_{4i}$, $T_{2i} = 0.6T^*_{2i} + 0.4T_{4i}$. For the third and fourth levels the coefficients become 0.4, 0.6, and 0.2, 0.8 respectively. Thus, level one represents the smallest departure from the null hypothesis and level four represents the largest. In particular, the Pearson partial correlation for level one is 0.06, and for levels two, three and four, it becomes 0.31, 0.69, and 0.94 respectively. From 1000 generated data sets from each of the four levels the null hypothesis was rejected 109, 345, 812, and 1000 times, for levels one, two, three, and four, respectively.

Recall that we chose $\alpha = 0.05$. Thus, if the small-sample distribution of the test statistic is well approximated by its asymptotic distribution, the attained level of the test procedure should be approximately 0.05 (i.e. it should reject about 50 times out of 1000 simulations under the null hypothesis). Large deviations from that indicate poor approximation to the small-sample distribution. From the statistical point of view, the attained level of 0.071 is significantly different from the chosen $\alpha = 0.05$, (a 95% confidence interval for the attained level is $(0.055, 0.087)$). From the practical point of view, however, the difference is not significant; in fact, for a sample of size 30 with 20% censoring, it is quite satisfactory. The power of the procedure is rather low for small departures from the null hypothesis but it increases very noticeably as the departures become more pronounced. For larger sample sizes, the attained level should be closer to 0.05, and the power should be greater. To verify this we ran again the simulations changing only the sample size to 80. With this sample size, the attained level was 0.049, and the power against the four alternatives was 0.158, 0.746, 0.999, and 1.000.

### 3.2 Application to astronomical data

As an application of the procedure described in §2 to an astrophysical problem we further investigated the sample already discussed by Morganti et al. (1993), Tadhunter et al. (1993) and Siebert et al. (1995). In total it consists of 88 sources (68 radio galaxies, 18 quasars, 2 BL Lac objects) which were selected from the Wall & Peacock 2.7 GHz sample (Wall & Peacock 1985) of radio sources. The selection criteria were: redshift z $< 0.7$, radio flux density $S_{2.7GHz} > 2$ Jy and declination $\delta < 10°$.

One of the key issues of the study was to investigate the relationship of the radio to the soft X–ray emission in the (0.1–2.4)keV ROSAT energy band. In Figures 1 and 2 we show a plot of the soft X-ray luminosity $L_x$ versus the total radio luminosity $L_t$ and the core radio luminosity $L_c$ , respectively. Clearly, a correlation is visible in both diagrams. Indeed, the correlation and regression analysis using ASURV (La Valley et al. 1992, Feigelson & Nelson 1985, Isobe et al. 1986) shows that the radio and the soft X–ray emission are correlated, both for the galaxies and the quasars, although the statistical significances of the correlations are low in the case of the quasars. This is probably due to the small sample size and the small range in luminosity.

Because of the flux limit of the original Wall & Peacock radio catalog, $L_t$ is strongly correlated with redshift. Further, the correlations of $L_x$ with $L_c$ and $L_t$ are not mutually independent since $L_c$ is also correlated with $L_t$ . In order to evaluate the influence of the individual redshift–luminosity correlations and the $L_c$ - $L_t$ correlation on the correlations

**Table 1.** Results of the correlation and regression analysis

| | N | X<br>$N_{UL}$ | Y<br>$N_{UL}$ | $\tau_{rx}$<br>$P_\tau$ | $\tau_{rx,z}$<br>$\sigma$ | P | $\tau_{rx,L_c}$<br>$\sigma$ | P |
|---|---|---|---|---|---|---|---|---|
| (1) | (2) | (3) | (4) | (5) | (6) | (7) | (8) | (9) |
| quasars | 18 | $\log L_{r,total}$ | $\log L_x$ | 0.250 | 0.196 | 0.271 | 0.036 | 0.749 |
| | | 0 | 1 | 0.081 | 0.179 | | 0.111 | |
| | 17 | $\log L_{r,core}$ | $\log L_x$ | 0.309 | 0.269 | 0.017 | 0.190 | 0.039 |
| | | 0 | 0 | 0.017 | 0.173 | | 0.127 | |
| galaxies | 68 | $\log L_{r,total}$ | $\log L_x$ | 0.264 | 0.115 | 0.075 | 0.185 | 0.0004 |
| | | 0 | 28 | 0.0003 | 0.065 | | 0.065 | |
| | 59 | $\log L_{r,core}$ | $\log L_x$ | 0.311 | 0.254 | $6.3 \times 10^{-5}$ | 0.249 | $2.7 \times 10^{-5}$ |
| | | 10 | 20 | $< 10^{-6}$ | 0.059 | | 0.060 | |

*Notes. Column (1):* AGN class. *Column (2):* Number of objects in each class. *Column (3),(4):* Independent(X) and dependent (Y) variable respectively. The number of upper limits is given in the second line. *Column (5):* Kendall's $\tau$ of the radio vs X-ray correlation with the corresponding probability that the correlation arises by chance given in the second line. *Column (6):* Partial Kendall's $\tau$ with the effect of redshift excluded, together with the calculated variance (see §2). *Column (7):* Probability of erroneously rejecting the null hypothesis (i.e. no correlation). *Column (8),(9):* Same as in columns (6) and (7), but with the effect of the $L_c - L_t$ correlation taken into account.

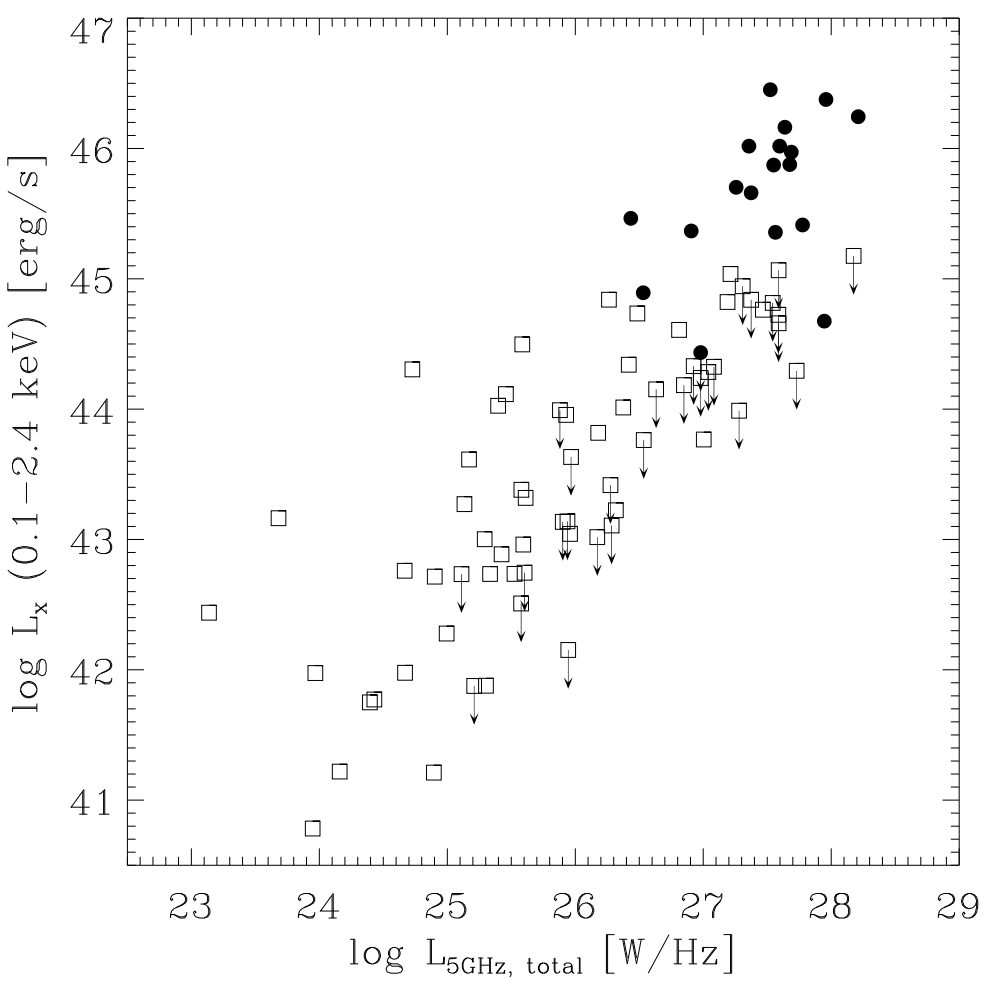

**Figure 1.** Total rest frame 2.7GHz radio luminosity versus soft X–ray luminosity in the (0.1–2.4)keV energy band. Full dots denote quasars, whereas galaxies are plotted with open squares. Upper limits are indicated by arrows.

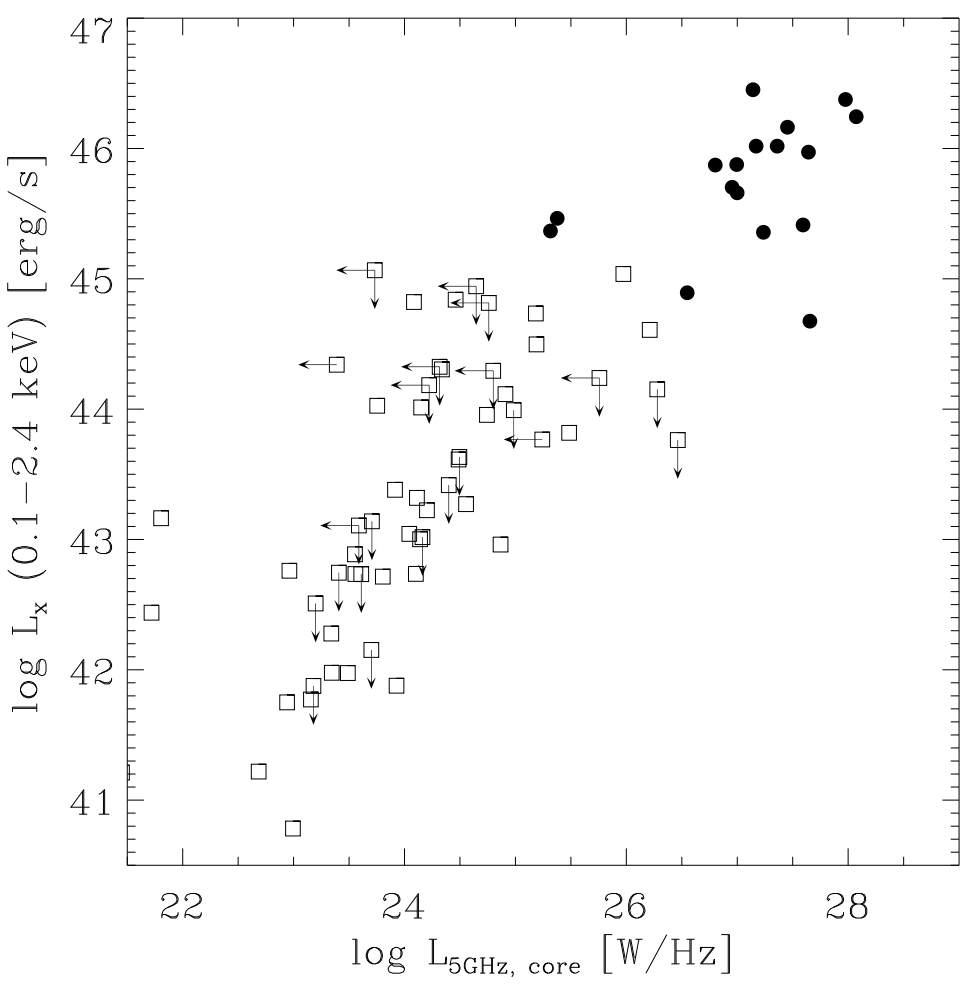

**Figure 2.** Radio core luminosity at 2.7 GHz versus soft X–ray luminosity. Full dots denote quasars, whereas galaxies are plotted with open squares. Upper limits are indicated by arrows.

with $L_x$ , we applied the procedure developed in §2 to this data set.

In the case of the quasars, the $L_x - L_t$ correlation seems to be strongly affected by both the redshift bias and the $L_c - L_t$ correlation. It turns out that the correlation is no longer statistically significant once both selection effects are properly accounted for. The $L_x - L_c$ correlation is much less affected and the probability of erroneously rejecting the null hypothesis of no correlation is $\lesssim 4\%$. As we have shown in the previous section, the power of the statistical test depends on the sample size. Given the low number of quasars, an error probability of 4% is acceptable. We thus conclude that there is indeed a correlation between $L_x$ and $L_c$ for quasars and that the $L_x - L_t$ correlation is probably an artifact of the redshift bias and/or the strong relation of $L_t$ with $L_c$ .

The results for the radio galaxies are similar. The $L_x$ vs $L_c$ correlation remains highly significant in the partial correlation analysis, whereas we find evidence that the $L_x - L_t$ correlation is most likely introduced by the redshift bias.

The fact that the $L_x$ - $L_c$ correlation is independent of redshift effects in both object classes is not surprising, since, because of the inclusion of upper limit values in the analysis, $L_x$ as well as $L_c$ do not depend *a priori* on redshift.

For a discussion of the results with respect to unification schemes and physical emission processes, see Siebert et al. (1995).

## 4 SUMMARY

In this paper we present a new methodology to test for partial association in censored (astronomical) data. This procedure is based on the Kendall's $\tau$ statistic and allows for the first time to assign a significance level to the resulting partial correlation coefficient. Tests with simulated data show that the procedure gives reliable results, although the power of the statistical test also depends on the sample size.

We applied the new method to a sample of 18 quasars and 68 radio galaxies defined in Morganti et al. (1993) in order to investigate the influence of two selection effects on the observed correlation of $L_x$ with both $L_t$ and $L_c$ , namely the strong correlations of $L_t$ with redshift and with $L_c$ . Whereas we find evidence that the $L_x$ - $L_t$ correlation is most likely an artifact of the redshift bias in both object classes, we conclude that the $L_x - L_c$ correlation is not affected by either of the selection effects in galaxies as well as in quasars.

## ACKNOWLEDGMENTS

The work of MA was supported in part by NSF grant DMS-9208066. MA thanks Tom Hettmansperger for a discussion about the statistical state of the art for the partial Kendall's $\tau$. JS thanks his colleagues from the ROSAT group for their support and Eric Feigelson for pointing out to him the existence of the SCCA.

## APPENDIX A: MATHEMATICAL DERIVATIONS

The idea is to express the numerator of Kendall's partial $\tau$ as a $U$-statistic and then use existing theory (Lee (1990); Serfling (1980)). We will use the notation introduced in Section 2. Write

$$
\begin{aligned}
\hat\tau_{12} \quad - \quad & \hat\tau_{13}\hat\tau_{23} \\
= \; & \frac{2}{n(n-1)}\sum_{i<j} h_{12}(i,j) - \\
& -\frac{4}{n^2(n-1)^2}\sum_{i<j} h_{13}(i,j)\sum_{i<j} h_{23}(i,j) \\
= \; & \frac{4}{n^2(n-1)^2}\sum_{i<j}\sum_{i_1<j_1}\left[h_{12}(i,j) - h_{13}(i,j)h_{23}(i_1,j_1)\right] \\
= \; & \frac{1}{n^2(n-1)^2}\sum_{i\neq j}\sum_{i_1\neq j_1}\tilde g(i,j,i_1,j_1) \\
= \; & \frac{1}{n(n-1)(n-2)(n-3)}\sum_{i\neq j\neq i_1\neq j_1}\tilde g(i,j,i_1,j_1) + \\
& +O(\frac{1}{n}) \\
= \; & \frac{1}{n(n-1)(n-2)(n-3)}\sum_{i\neq j\neq i_1\neq j_1} g(i,j,i_1,j_1) + \\
& +O(\frac{1}{n}) \\
= \; & \frac{24}{n(n-1)(n-2)(n-3)}\sum_{i<j<i_1<j_1} g(i,j,i_1,j_1) + \\
& +O(\frac{1}{n})
\end{aligned}
$$

where $O(\frac{1}{n})$ denotes a quantity that when multiplied by $n$ remains bounded as $n \to \infty$. The first term on the right hand side is a $U$-statistic, that has mean value zero under the null hypothesis. Thus, from Serfling (1980, p. 188) it follows that, under the null hypothesis, $\hat\tau_{12} - \hat\tau_{13}\hat\tau_{23}$ has the same asymptotic distribution as its 'projection'

$$\frac{4}{n}\sum_{i=1}^{n} P_i,$$

where the $P_i$ are independent and identically distributed random variables and are described in the preceding reference. Thus its asymptotic variance is $16n^{-1}\mathrm{Var}(P_1)$. The estimate of $\mathrm{Var}(P_1)$ given in (1) is the estimate proposed by Sen (1960) modified to increase the sensitivity of the testing procedure under the alternative hypothesis.